\documentclass[twoside,twocolumn,9pt]{article}
\usepackage{extsizes}
\usepackage[super,sort&compress,comma]{natbib} 
\usepackage[version=3]{mhchem}
\usepackage[left=1.5cm, right=1.5cm, top=1.785cm, bottom=2.0cm]{geometry}
\usepackage{balance}
\usepackage{mathptmx}
\usepackage{sectsty}
\usepackage{graphicx} 
\usepackage{lastpage}
\usepackage[format=plain,justification=justified,singlelinecheck=false,font={stretch=1.125,small,sf},labelfont=bf,labelsep=space]{caption}
\usepackage{float}
\usepackage{fancyhdr}
\usepackage{fnpos}
\usepackage[english]{babel}
\addto{\captionsenglish}{%
  \renewcommand{\refname}{Notes and references}
}
\usepackage{array}
\usepackage{droidsans}
\usepackage{charter}
\usepackage[T1]{fontenc}
\usepackage[usenames,dvipsnames]{xcolor}
\usepackage{setspace}
\usepackage[compact]{titlesec}
\usepackage{hyperref}

\usepackage{epstopdf}
\usepackage{subfig}

\definecolor{cream}{RGB}{222,217,201}

\begin{document}

\pagestyle{fancy}
\thispagestyle{plain}
\fancypagestyle{plain}{
\renewcommand{\headrulewidth}{0pt}
}

\makeFNbottom
\makeatletter
\renewcommand\LARGE{\@setfontsize\LARGE{15pt}{17}}
\renewcommand\Large{\@setfontsize\Large{12pt}{14}}
\renewcommand\large{\@setfontsize\large{10pt}{12}}
\renewcommand\footnotesize{\@setfontsize\footnotesize{7pt}{10}}
\makeatother

\renewcommand{\thefootnote}{\fnsymbol{footnote}}
\renewcommand\footnoterule{\vspace*{1pt}%
\color{cream}\hrule width 3.5in height 0.4pt \color{black}\vspace*{5pt}} 
\setcounter{secnumdepth}{5}

\makeatletter 
\renewcommand\@biblabel[1]{#1}            
\renewcommand\@makefntext[1]%
{\noindent\makebox[0pt][r]{\@thefnmark\,}#1}
\makeatother 
\renewcommand{\figurename}{\small{Fig.}~}
\sectionfont{\sffamily\Large}
\subsectionfont{\normalsize}
\subsubsectionfont{\bf}
\setstretch{1.125} 
\setlength{\skip\footins}{0.8cm}
\setlength{\footnotesep}{0.25cm}
\setlength{\jot}{10pt}
\titlespacing*{\section}{0pt}{4pt}{4pt}
\titlespacing*{\subsection}{0pt}{15pt}{1pt}

\fancyfoot{}
\fancyfoot[LO,RE]{\vspace{-7.1pt}\includegraphics[height=9pt]{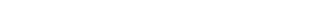}}
\fancyfoot[CO]{\vspace{-7.1pt}\hspace{11.9cm}\includegraphics{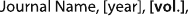}}
\fancyfoot[CE]{\vspace{-7.2pt}\hspace{-13.2cm}\includegraphics{head_foot/RF}}
\fancyfoot[RO]{\footnotesize{\sffamily{1--\pageref{LastPage} ~\textbar  \hspace{2pt}\thepage}}}
\fancyfoot[LE]{\footnotesize{\sffamily{\thepage~\textbar\hspace{4.65cm} 1--\pageref{LastPage}}}}
\fancyhead{}
\renewcommand{\headrulewidth}{0pt} 
\renewcommand{\footrulewidth}{0pt}
\setlength{\arrayrulewidth}{1pt}
\setlength{\columnsep}{6.5mm}
\setlength\bibsep{1pt}

\makeatletter 
\newlength{\figrulesep} 
\setlength{\figrulesep}{0.5\textfloatsep} 

\newcommand{\topfigrule}{\vspace*{-1pt}%
\noindent{\color{cream}\rule[-\figrulesep]{\columnwidth}{1.5pt}} }

\newcommand{\botfigrule}{\vspace*{-2pt}%
\noindent{\color{cream}\rule[\figrulesep]{\columnwidth}{1.5pt}} }

\newcommand{\dblfigrule}{\vspace*{-1pt}%
\noindent{\color{cream}\rule[-\figrulesep]{\textwidth}{1.5pt}} }

\makeatother

\twocolumn[
  \begin{@twocolumnfalse}
{\includegraphics[height=30pt]{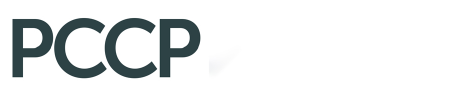}\hfill\raisebox{0pt}[0pt][0pt]{\includegraphics[height=55pt]{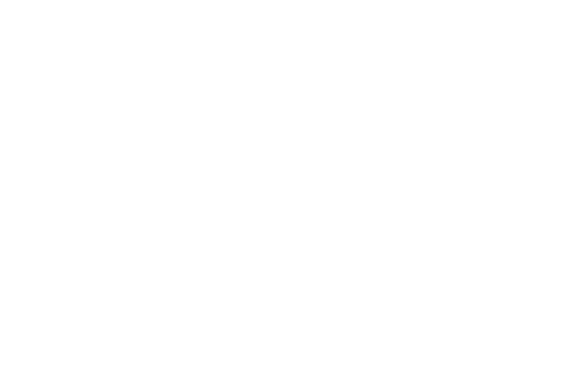}}\\[1ex]
\includegraphics[width=18.5cm]{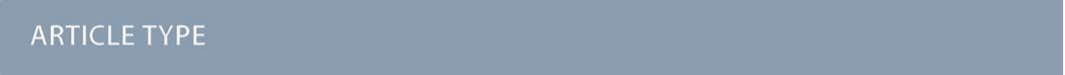}}\par
\vspace{1em}
\sffamily
\begin{tabular}{m{4.5cm} p{13.5cm} }

\includegraphics{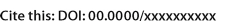} & \noindent\LARGE{\textbf{Influence of Three-Body Effects on Halogen Bonding$^{\dag \ddag}$}} \\
\vspace{0.3cm} & \vspace{0.3cm} \\

 & \noindent\large{Sharon A. Ochieng\textit{$^{a}$} and Konrad Patkowski$^\ast$\textit{$^{a}$}} \\

\includegraphics{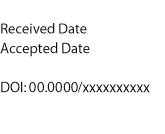} & \noindent\normalsize{We construct a new noncovalent benchmark dataset 3BXB that combines halogen-bonded bimolecular complexes from the SH250 dataset [K\v{r}\'i\v{z} and \v{R}ez\'a\v{c}, \textit{Phys. Chem. Chem. Phys.}, 2022, \textbf{24}, 14794] with a third interacting partner, either H$_2$O or CH$_4$. The reference total and three-body interaction energies are computed at the CCSD(T) level. To shed light on the physical origins of binding and cooperativity in complexes of this kind, several symmetry-adapted perturbation theory (SAPT)-based energy decompositions were performed for both pairwise additive and nonadditive terms. We found that the two-body attractions in the 3BXB complexes are dominated by either electrostatics or dispersion, while the three-body effect is dominated by induction and can be either attractive or repulsive. An accurate recovery of reference interaction energies is attained by the wavefunction-based two-body SAPT variants including the $\delta$MP2 correction, combined with the SAPT(DFT) estimates of nonadditive induction and first-order exchange and {\em any} estimate of nonadditive dispersion. The values for the latter term are sometimes quite inconsistent between different approaches; fortunately, nonadditive dispersion is a relatively minor effect for complexes studied here, and all reasonable estimates lead to total interaction energies of similar accuracy.} \\

\end{tabular}

 \end{@twocolumnfalse} \vspace{0.6cm}

  ]

\renewcommand*\rmdefault{bch}\normalfont\upshape
\rmfamily
\section*{}
\vspace{-1cm}


\footnotetext{\textit{$^{a}$~Department of Chemistry and Biochemistry, Auburn University, Auburn, Alabama 36849, United States}
}
\footnotetext{$\ast$~Corresponding author, email: patkowsk@auburn.edu.}


\footnotetext{\dag~Supplementary Information available: [details of any supplementary information available should be included here]. See DOI: 10.1039/cXCP00000x/}
\footnotetext{\ddag~Dedicated to Professor Giuseppe Resnati, celebrating a career in fluorine and noncovalent chemistry on the occasion of his 70th birthday.}



\section{Introduction}\label{sec:intro}
Accurate computational data are vital in advancing chemical understanding in domains such as template-directed synthesis, enzyme–ligand binding mechanisms, and the structure and properties of materials. The significance of computations cannot be overemphasized, because many molecular and electronic properties responsible for a broad range of phenomena cannot be measured directly.  Benchmarks, therefore, play a vital role in validation of the accuracy of approximate computational methods and the development of those that rely on empirical parameters. In the realm of noncovalent interactions, where experimental data does not typically offer a straightforward comparison, these datasets are vital for ensuring accuracy and facilitating method development. Many datasets addressing different kinds of noncovalent interactions have been introduced. 
They range from well established but relatively small sets of complexes such as S22\cite{Jurecka2006BenchmarkPairs}, 
S66\cite{Rezac2011S66:Structures},
X40\cite{Rezac2012BenchmarkMolecules}, or
S12L\cite{Grimme2012SupramolecularTheory},
to more extensive sets such as SSI\cite{Burns2017TheInteractions} (interactions between amino acid side chains in proteins), NENCI\cite{Sparrow2021NENCI-2021.Contacts} (a wide variety of complexes in radially and orientationally nonequilibrium configurations),  or  the NCI Atlas by \v{R}ez\'a\v{c} et al. with several subsets mapping pairwise interactions over broad chemical space one binding type at a time\cite{Rezac2020Non-CovalentBonding,Rezac2020Non-CovalentSpace,Kriz2021Non-CovalentContacts,Kriz2022Non-covalentInteractions,Rezac2022Non-CovalentSpace}. 
 

In molecular clusters, liquids, and solids, interactions between pairs of molecules provide only an approximate description of the total binding energy.
Nonadditive contributions from triplets (and sometimes even quadruplets) of molecules are essential for obtaining the correct energetic order of water hexamers \cite{Gora2011InteractionExpansion} and energetically ranking different polymorphs of molecular crystals \cite{Beran2016ModelingTheory}.
However, the existing benchmark datasets for molecular trimers (or larger clusters) are not nearly as broad and are dominated by data for water clusters \cite{Temelso2011BenchmarkCorrections} 
or for clusters involving an ion such as Cl$^{-}$ solvated by a few water molecules\cite{Lao2015AccurateMethods}.
Only a handful of benchmark three-body datasets contains trimers of different chemical character.
The 3B-69 dataset contains 69 (A)$_3$ trimers  from 23 different molecular crystals \cite{Rezac2015BenchmarkMethods}.
The S22(3) set \cite{Alkan2019Many-BodyClusters} includes 24 (A)$_3$ and (A)$_2$B-type trimers obtained by augmenting the complexes in the popular S22 database \cite{Jurecka2006BenchmarkPairs}.
Finally, the 3BHET dataset from our earlier work \cite{Ochieng2023AccurateDecomposition} comprises 20 model (A)$_2$B- and ABC-type trimers exhibiting diverse types of interactions.
However, these are relatively small datasets built for specific purposes and only focus on fairly specific types of interactions. Therefore, the field would benefit from the development of benchmark nonadditive datasets that are larger, more general, and as accurate as possible.    

In this work, we build on the SH250\cite{Kriz2022Non-covalentInteractions} dataset of halogen bonded complexes featuring Cl, Br, I, and extend it to trimers to study the effects of three-body interactions on the halogen bond (XB).  In recent years, noncovalent complexes involving $\sigma$-hole interactions, including halogen  and chalcogen bonding, have gained significant attention \cite{Cavallo2016TheBond,Haberhauer2020TheHeterocycles,Mahmudov2022ChalcogenChemistry}.
These interactions are characterized by attractive forces between an electrophilic region on a halogen or chalcogen atom and a Lewis base. This is as a result of an anisotropic redistribution of the halogen electron density. The positive region that appears at the extension of a $\sigma$ covalent bond is termed a “$\sigma$-hole”\cite{Stone2013AreDriven,Davis2017HarnessingReactions,Politzer2017TheRevisited}.
Hydrogen and halogen bonding act as complementary design features for the design of functional molecules and materials, stabilization of drug-protein complexes, and self-assembly of supramolecular structures\cite{Metrangolo2005HalogenBonding,Baldrighi2013HalogenPreservative,Bertolani2015SupramolecularIodination}.
Consequently, the development of robust computational methods to describe halogen bonding is widely important. Ideal methods not only aid in accurately modeling and understanding these interactions, but also ensure reliability in predicting interaction energies for new systems. Additionally, when the accurate methods become prohibitively costly, it is common practice to test the interaction energies from approximate theoretical methods against accurate benchmark calculations.

The extension of the halogen bonded part of the SH250 dataset to trimers presented here aims to investigate the interplay between halogen bonding and other interactions.
Besides introducing the dataset, this work also focuses on its applications to benchmarking different variants of symmetry-adapted perturbation theory (SAPT). This study was motivated by our previous work on heteromolecular trimers which observed that the XB complexes exhibited a comparatively smaller magnitude of the nonadditive exchange and dispersion contributions from a SAPT(DFT) energy decomposition analysis\cite{Ochieng2023AccurateDecomposition}. With only three systems from a dataset with 20 trimers, that observation was possibly accidental as the data were too small to be useful in making predictions on general trends in specific interactions such as XB. 
Building on these findings, the focus on trimers featuring XB in this work seeks to attain a broader coverage with a more comprehensive benchmark of trimers in a larger chemical space.


\section{Computational Methods}\label{sec:methods}
\subsection{Composition of the data set and benchmark interaction energies}

\begin{figure*}

\subfloat[\tiny 1 chlorobenzene--trimethylamine--water]{\includegraphics[width=0.28\textwidth]{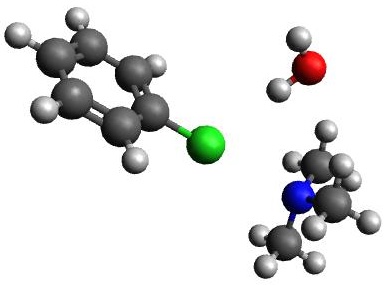}}\label{fig:f_min_saptdft_nonadd_water}
\hfill
\subfloat[\tiny 2 chlorine--pyridine-N-oxide--methane]{\includegraphics[width=0.28\textwidth]{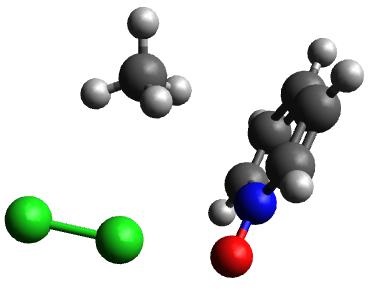}}\label{fig:f_min_saptdft_nonadd_methane}
\hfill
\subfloat[\tiny 3 bromine--trimethylamine--water]{\includegraphics[width=0.23\textwidth]{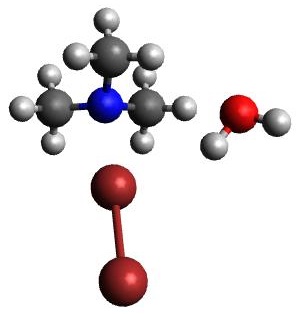}}\label{fig:f_max_Eint_water}
\hfill
\subfloat[\tiny 4 bromine--trimethylamine--methane]{\includegraphics[width=0.28\textwidth]{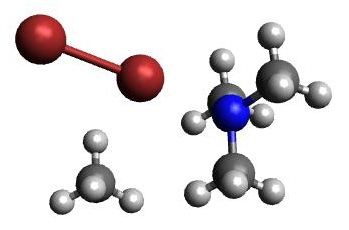}}\label{fig:f_max_Eint_methane}
\hfill
\subfloat[\tiny 5 bromine--thioacetone--methane]{\includegraphics[width=0.28\textwidth]{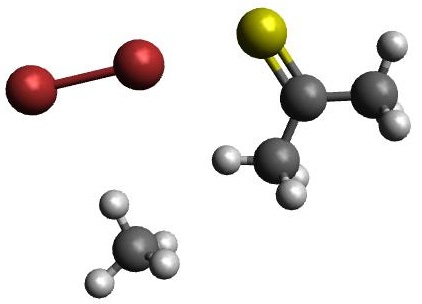}}\label{fig:f_max_saptdft_nonadd_methane}
\hfill
\subfloat[\tiny 6 trifluorobromomethane--chlorine--water]{\includegraphics[width=0.2\textwidth]{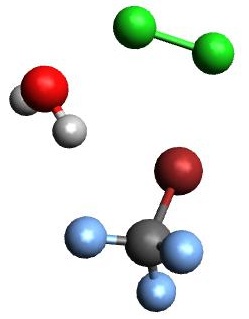}}\label{fig:f_min_E_int_water}
\hfill
\subfloat[\tiny 7 iodine--pyridine-N-oxide--water]{\includegraphics[width=0.28\textwidth]{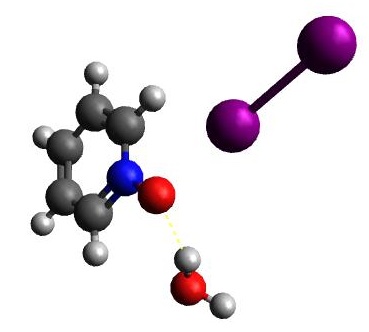}}\label{fig:f_min_E3int_nonadd_water}
\hfill
\subfloat[\tiny 8 trifluoroiodomethane--pyridine-N-oxide--methane]{\includegraphics[width=0.25\textwidth]{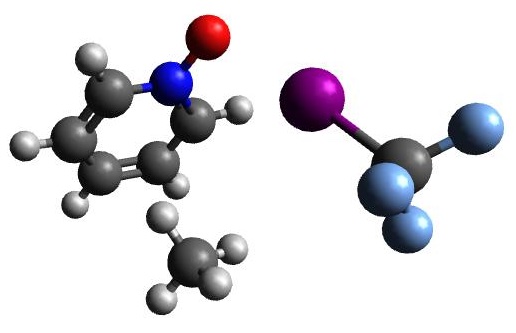}}\label{fig:f_min_E3int_nonadd_methane}
\hfill
\subfloat[\tiny 9 trifluoroiodomethane--fluorine--methane]{\includegraphics[width=0.28\textwidth]{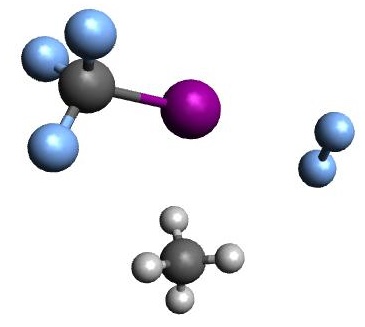}}\label{fig:f_min_Eint_nonadd_methane}
\hfill
\subfloat[\noindent \tiny 10 chlorine--thioacetone--water]{\includegraphics[width=0.2\textwidth]{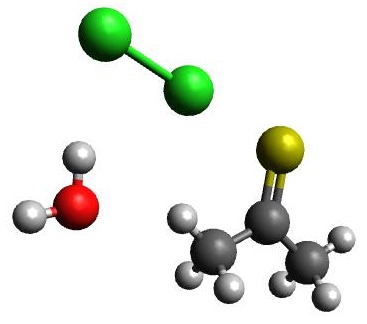}}\label{fig:f_max_E3int_and_saptdft_nonadd_water}
\hfill
\subfloat[\tiny 11 chlorine--bromine--water]{\includegraphics[width=0.2\textwidth]{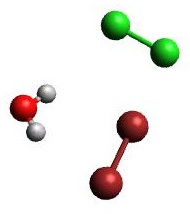}}\label{fig:f_elst_cl2_br2_h2o}
\hfill
\subfloat[\tiny 12 chlorine--iodine--water]{\includegraphics[width=0.18\textwidth]{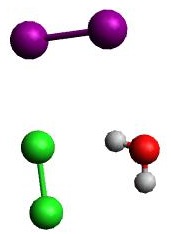}}\label{fig:f_elst_cl2_i2_h2o}
\hfill
\caption{Representative trimer structures from the 3BXB dataset.}\label{fig:struct1}
\end{figure*}

The dataset introduced in this work, which will be denoted 3BXB, follows the same strategy as our earlier 3BHET dataset\cite{Ochieng2023AccurateDecomposition} but with a focus on halogen bonding. 
We examine the influence of the three-body non-additive interactions on the halogen bond, looking for common trends in the three-body contributions that might characterize those halogen-bonded structures. 
3BXB comprises 214 trimers, 107 with water and 107 with methane, obtained by extending the halogen-bonded dimers featuring Cl, Br and I from the SH250 dataset\cite{Kriz2022Non-covalentInteractions}.
We picked either water or methane as the third partner because the two molecules have the same number of electrons (10), but exhibit very different kinds of interactions, predominantly electrostatic/hydrogen bonded for water and mostly dispersive for methane. 
The starting geometries were obtained from dimers in the SH250\cite{Kriz2022Non-covalentInteractions} dataset by adding the third partner and reoptimizing only the variables pertaining to the third molecule in the trimer. This ensures that the halogen-bonded complex is not broken, but simply augmented with an external partner. 
The structures of some of these trimers are illustrated in Fig.~\ref{fig:struct1}, and the Cartesian coordinates for all 214 trimers are provided in the Supporting Information.
All restricted geometry optimizations were carried out using the MP2 method and the aug-cc-pVDZ basis set. Frequency calculations (in an appropriate subspace of parameters) were also carried out to confirm that a minimum has been reached in that subspace.

\captionsetup[subfigure]{labelformat=empty}

To ensure a seamless comparison with the previous 3BHET dataset \cite{Ochieng2023AccurateDecomposition}, the benchmark interaction energy calculations are performed using the same composite CCSD(T)/CBS scheme. 
The interaction energies of interest are the two-body and nonadditive three-body terms, where the latter is defined as the difference between the trimer interaction energy and the sum of the pair interaction energies\cite{Patkowski2020RecentTheory}. Specifically, the complete interaction energy in the trimer
\begin{equation}
E_{int} = E_{ABC} - E_{A} - E_{B} - E_{C} 
\end{equation}
can be decomposed as
\begin{equation}
E_{int} = E^2_{int} + E^3_{int} 
\end{equation}
where the two-body part is the sum of pairwise interactions:
\begin{equation}
E^2_{int} = (E_{AB} - E_A - E_B) + (E_{BC} - E_B - E_C) + (E_{AC} -E_A - E_C) 
\end{equation}
The monomer and dimer calculations were performed in the full trimer basis set to eliminate basis set superposition error (BSSE). In addition, the interaction energies were computed using fixed monomer geometries ignoring all monomer deformation effects\cite{Rezac2016BenchmarkApplications}.
The benchmark interaction energies at both the two-body and three-body level were calculated using the composite CCSD(T)/CBS approach where correlation consistent basis sets fully augmented with diffuse functions were used; in this text, aug-cc-pVXZ (X = D, T, Q, 5) are abbreviated as aXZ.
The benchmark calculations employ the well-kown composite CCSD(T)/CBS scheme that has contributions of the Hartree-Fock method (HF), the MP2 correlation energy, and a small-basis  CCSD(T) correction: 
\begin{equation}\label{eq:composite}
E^{\rm (CCSD(T)/CBS)} = E^{\rm HF} + E_{\rm corr}^{\rm MP2/CBS} + {{\delta}E^{\rm CCSD(T)}}    
\end{equation}

In this formalism, a single HF calculation in a large basis set, a5Z basis for this purpose, is sufficient as it converges fast with the basis set size. The MP2 correlation energy is extrapolated to the CBS limit from the aQZ and a5Z basis sets using Helgaker's $X^{-3}$ formula \cite{Halkier1998Basis-setO} and the coupled cluster correction $\delta E^{CCSD(T)}$ is calculated in the aDZ basis set.
Thus, the benchmark interaction energies produced in this work correspond to the so-called silver standard of theory \cite{Burns2014AppointingApproaches,Kodrycka2019PlatinumCalculations}.

\subsection{Two-body SAPT interaction energy decomposition}

For appropriate interpretation of the supermolecular data obtained from the benchmark calculations, we need to know the nature of both the two- and three-body binding energies in each trimer. This is efficiently achieved through energy decomposition analysis such as SAPT.  In this study, we employ several SAPT formalisms \cite{Jeziorski1994PerturbationComplexes, Parker2014LevelsEnergies,Patkowski2020RecentTheory}
to decompose the interaction energy into physically meaningful components. While we utilize various flavors of SAPT, the formalisms described below follow a set of general rules that apply across these methods. These rules provide a consistent framework for interpreting the interaction energies and their components, regardless of the specific SAPT variant employed.

In the two-body and three-body SAPT based on the HF wavefunctions, the Hamiltonian is partitioned into the Fock operator $F$, the intermolecular interaction operator $V$, and the intramolecular correlation operator $W$ encompassing all monomers:  

\begin{equation}
  H = F + V + W  
\end{equation}
where (in the three-body case) \begin{equation}
    F = F_A + F_B + F_C\;\;\;\;\;\;
    V = V_{AB} + V_{BC} + V_{AC}\;\;\;\;\;\;
    W = W_A + W_B + W_C
\end{equation} 
The simplest approximation that provides qualitatively reasonable total interaction energies is the SAPT0 one, which totally neglects the intramonomer electron correlation\cite{Patkowski2020RecentTheory}. 
The two-body SAPT0 interaction energy is calculated as
\begin{align}
\label{eq:sapt0}
    E^{\rm SAPT0}_{int} = & E^{(10)}_{elst} + E^{(10)}_{exch} + E^{(20)}_{ind,resp} + E^{(20)}_{exch-ind,resp} + E^{(20)}_{disp} + E^{(20)}_{exch-disp} 
    \nonumber \\ & + \delta{E^{(2)}_{HF}}
\end{align} 

In a SAPT correction $E^{(kl)}$ in Eq.~(\ref{eq:sapt0}) and throughout the text, the subscripts $k$ and $l$ denote orders of perturbation theory with respect to $V$ and $W$, respectively.
SAPT0 provides the physical components that represent the electrostatic, first-order exchange, induction, exchange-induction (both with the inclusion of monomer relaxation/response effects as denoted by the additional subscript ``resp''), dispersion, exchange-dispersion, and $\delta{E^{(2)}_{HF}}$ contributions. 
The $\delta{E^{(2)}_{HF}}$ term is meant to estimate the higher-order induction and exchange-induction effects from a supermolecular HF calculation\cite{Parker2014LevelsEnergies,Patkowski2020RecentTheory}:
\begin{equation}
\delta E^{\rm HF}_{int} = E^{\rm HF}_{int}  -E^{(10)}_{elst}- E^{(10)}_{exch} - E^{(20)}_{ind,resp} - E^{(20)}_{exch-ind,resp}
\label{eq:2bdelta}
\end{equation}
This energy decomposition allows one to classify complexes according to the relative importance of electrostatics, induction, and dispersion for their binding, for example, using a ternary diagram\cite{Smith2016RevisedTheory}. 

To attain quantitative accuracy of total energies and their contributions, a higher-order SAPT level that includes both intra- and intermonomer electron correlation effects is required. 
In this study, the two-body interaction energy decomposition was performed by wavefunction-based SAPT up to the SAPT2+3 level of theory, 
simultaneously including the effects of the $W$ operator on most SAPT0-level corrections  and some terms that are third-order in $V$. Beyond SAPT0, the following levels of SAPT have been established: \cite{Parker2014LevelsEnergies}
\begin{align}
E^{\rm SAPT2}_{int} =  & E^{(10)}_{elst} + E^{(12)}_{elst,resp} \nonumber \\ & + E^{(10)}_{exch} + E^{(11)}_{exch} + E^{(12)}_{exch} \nonumber \\ & +  E^{(20)}_{ind,resp} + E^{(20)}_{exch-ind,resp} + E^{(22)}_{ind} + E^{(22)}_{exch-ind} + \delta{E^{(2)}_{HF}} \nonumber \\ & 
+ E^{(20)}_{disp} + E^{(20)}_{exch-disp} 
\label{eq:SAPT2}
\end{align}

\begin{align}
E^{\rm SAPT2+}_{int} = E^{\rm SAPT2}_{int} + E^{(21)}_{disp} + E^{(22)}_{disp}  
\end{align}

\begin{align}
E^{\rm SAPT2+(3)}_{int} = E^{\rm SAPT2+}_{int} + E^{(13)}_{elst,resp} + E^{(30)}_{disp} 
\end{align}
\begin{align}
E^{\rm SAPT2+3}_{int} = & E^{\rm SAPT2+(3)}_{int}
 + [ \delta{E^{(3)}_{HF}} - \delta{E^{(2)}_{HF}} + E^{30}_{ind,resp} + E^{30}_{exch-ind,resp}] 
\nonumber \\ & +[E^{(30)}_{exch-disp} + E^{(30)}_{ind-disp} + E^{(30)}_{exch-ind-disp}]
\label{eq:SAPT2+3}
\end{align}

Equations (\ref{eq:SAPT2})--(\ref{eq:SAPT2+3}) show the progressive improvements to the SAPT0 approximation. 
In addition to those improvements, for levels of SAPT beyond SAPT2, we may include extra terms denoted $\delta E^{MP2}_{int}$ and (CCD). The $\delta E^{MP2}_{int}=E^{MP2}_{int}-E^{SAPT2}_{int}$ term is based on a supermolecular MP2 interaction energy and may account for higher-order couplings between induction and dispersion as well as for some exchange effects, including those beyond the commonly used single exchange approximation \cite{Parker2014LevelsEnergies}.
The (CCD) notation in turn signifies that the second-order dispersion energy has been computed nonperturbatively, using the CCD+ST(CCD) dispersion approach introduced by Williams et al.\cite{Williams1995DispersionExcitations} in place of the perturbative sum $E^{(20)}_{disp}+E^{(21)}_{disp}+E^{(22)}_{disp}$.

Alternative variants to the wavefunction-based SAPT approach, SAPT(DFT) and XSAPT, were also used to perform energy decompositions of both two- and three-body interaction energies. Here, we follow the same workflow as employed in our previous work for both SAPT(DFT) and XSAPT\cite{Ochieng2023AccurateDecomposition}.
For SAPT(DFT), the theory based on the Kohn-Sham (KS) description of the monomers, the Fock operator is replaced by the Kohn-Sham operator, $K= K_A + K_B$ or $K = K_A + K_B + K_C$ for dimers and trimers, respectively. Then, the two-body SAPT(DFT) interaction energy is computed as
\begin{eqnarray}
    E_{int}^{SAPT(DFT)} &=& E_{elst}^{(1)}(KS) + E_{exch}^{(1)}(KS) + E_{ind}^{(2)}(CKS)  + E_{exch-ind}^{(2)}(CKS) \nonumber \\ && + E_{disp}^{(2)}(CKS)+ E_{exch-disp}^{(2)}(CKS) +\delta E^{HF}_{int}
\end{eqnarray}
KS denotes standard uncoupled Kohn-Sham calculations (such SAPT(KS) quantities will also be used as components of the XSAPT approach below), CKS stands for the coupled Kohn-Sham, that is, the linear-response DFT approach to molecular frequency-dependent density susceptibilities (FDDS)\cite{Heelmann2005Density-functionalEnergies,Ochieng2023AccurateDecomposition}
which are then used to compute SAPT(DFT) induction (static FDDS) and dispersion (imaginary-frequency FDDS).
The exchange counterparts to dispersion and induction are also evaluated using the respective CKS amplitudes \cite{Heelmann2005Density-functionalEnergies} (that is, no scaling of the uncoupled quantities is performed), and they are computed using the single-exchange approximation prevalent in standard SAPT \cite{Jeziorski1994PerturbationComplexes}.
The  $\delta E^{\rm HF}_{int}$ term is once again taken from the HF-based SAPT theory, Eq.~(\ref{eq:2bdelta}). 
Since DFT fails to account for the correct asymptotic behavior of the exchange-correlation potential, the PBE0 functional used in all SAPT(DFT) computations was augmented with an asymptotic correction\cite{Misquitta2005Symmetry-adaptedMonomers}.

The extended SAPT (XSAPT) approach allows for calculations for an arbitrary number of monomers with all-body induction interactions taken into account in an iterative manner\cite{Lao2015AccurateMethods}. This goal is achieved through the explicit polarization (XPol) approach \cite{Xie2008TheField} that solves the self-consistent field equations for each fragment in an electrostatic embedding of all the other fragments in a form of point charges. The electron density obtained in this way is used to construct a new set of charges, and the process is iterated for all molecules until self-consistency is achieved. 
When combining SAPT with XPol, we obtain the total XSAPT interaction energy as 
\begin{align}\label{eq:XSAPT_E_Decomp}
    E^{\rm XSAPT}_{int} = &E^{(1)}_{elst} +  E^{(1)}_{exch} +  E^{(2)}_{disp+exch-disp} + [ E^{(2)}_{ind} +  E^{(2)}_{exch-ind} \nonumber\\ & + \sum_{A} \sum_{B>A} \delta E^{\rm HF}_{AB}  + \sum_{A} \sum_{B>A} ( E^{\rm XSAPT}_{AB} - E^{\rm SAPT(KS)}_{AB}) + E^{\rm MB}_{int}]
\end{align}
where the $E^{(1)}_{elst}$ and  $E^{(1)}_{exch}$ terms represent the sum of these energy components over all pairs of dimers, and the $E^{(2)}_{disp+exch-disp}$ term is modeled by a suitably adjusted many-body dispersion (MBD) approach \cite{Tkatchenko2012AccurateInteractions,Carter-Fenk2019AccurateDispersionc}. The term in square brackets represents the total induction energy, which includes a many-body contribution\cite{Lao2014Erratum:034107},  
and the $\delta E^{HF}$ term is assumed to be pairwise additive. 
The $E^{\rm XSAPT}_{AB}$ and $E^{\rm SAPT(KS)}_{AB}$ quantities are total SAPT(KS) energies for dimer AB computed with and without the XPol embedding, respectively, and $E^{\rm MB}_{int}$ denotes the pairwise-nonadditive effect of this embedding.
We neglected the additional three-body induction coupling terms that are optional in XSAPT \cite{Herbert2012RapidTheory}, as they are expensive to compute, and our previous study found that the effect of those couplings is small \cite{Ochieng2023AccurateDecomposition}.
Thus, XSAPT can take into account some nonadditive three-body (and higher-body) effects, namely induction (through the XPol iterative scheme) and dispersion (through the MBD formalism).
Note that the SAPT(KS) calculations underlying the XSAPT scheme are computed without an explicit asymptotic correction; instead, the long-range corrected functional LRC-$\omega$PBEh \cite{Vydrov2006AssessmentFunctional,Rohrdanz2009AStates} 
was used, with the range separation parameter $\omega$ tuned separately for each molecule using the global density-dependent (GDD) scheme \cite{Modrzejewski2013Density-dependentProperties}.

\subsection{Three-body SAPT energy decomposition and its alternatives}

The energy decomposition of nonadditive three-body effects was performed by either XSAPT or SAPT(DFT). While the XSAPT method was described in the previous section including all-body terms, here we focus on the three-body SAPT(DFT) approach \cite{Podeszwa2007Three-bodyMonomers}. 
In this case, the electrostatic energy is rigorously pairwise additive and so is the second-order dispersion energy; however, important nonadditivity effects do show up in the third-order dispersion energy.
The remaining corrections, most notably $E_{ind}^{(2)}$, are not pairwise additive and, in the three-body SAPT(DFT) framework, the nonadditive interaction energy is computed as \cite{Podeszwa2007Three-bodyMonomers}: 
\begin{eqnarray}
    E_{int,3}^{SAPT(DFT)} &=& \left(E_{exch,3}^{(1)}(KS)\right)_{exch} 
    \nonumber \\ && + \left(E_{ind,3}^{(2)}(CKS)  + \widetilde{E}_{exch-ind,3}^{(2)}(CKS) +\delta E^{HF}_{int,3}\right)_{ind}
     \nonumber \\ && + \left(\widetilde{E}_{exch-disp,3}^{(2)}(CKS) +  E_{disp,3}^{(3)}(CKS)\right)_{disp} 
\end{eqnarray}
where the grouping in parentheses defines the overall splitting of nonadditive effects into exchange, induction, and dispersion terms as utilized below.
As for the two-body energy, the first-order exchange is computed using the (asymptotically corrected) Kohn-Sham orbitals and orbital energies. 
In the second and third order, the induction and dispersion terms are computed from monomer FDDSs, and their exchange counterparts are approximated by scaling the respective uncoupled quantities as \cite{Podeszwa2007Three-bodyMonomers}
\begin{equation}\label{eq:scaleind2b}
\widetilde{E}^{(2)}_{exch-ind,3}(CKS)= E^{(2)}_{exch-ind,3}(KS) \frac{{E}^{(2)}_{ind,3}(CKS)}{{E}^{(2)}_{ind,3}(KS)}   
\end{equation}
\begin{equation}\label{eq:scaledisp2b}
\widetilde{E}^{(2)}_{exch-disp,3}(CKS) = E^{(2)}_{exch-disp,3}(KS) \frac{{E}^{(3)}_{disp,3}(CKS)}{{E}^{(3)}_{disp,3}(KS)}   
\end{equation}
For the exchange-induction term, the scaling factor was restricted to the $[-2.0;2.0]$ range to prevent overestimation when the uncoupled induction term is way smaller than the CKS term (which might accidentally happen as these contributions change sign between different trimer configurations). 
Finally, the nonadditive HF delta term is defined from the supermolecular HF nonadditive energy and its three-body SAPT0 approximation as
\begin{equation}
\delta E^{HF}_{int,3} = E^{HF}_{int,3} - E_{exch,3}^{(10)} - E_{ind,3}^{(20)} - E_{exch-ind,3}^{(20)}
\end{equation}

Among the  nonadditive components, the three-body dispersion is the most computationally demanding. Its many-body quantum nature 
makes it inherently difficult to calculate accurately. Previous studies have shown that the subtle three-body dispersion forces are overestimated by popular \textit{ab initio} quantum mechanical approximations such as DFT \cite{Tkatchenko2008PopularSystems} and they are completely missed by supermolecular MP2.
To reduce costs, it is common to adopt a simplified approximation and model three-body dispersion by its leading Axilrod-Teller-Muto (ATM) asymptotic term\cite{Axilrod1943InteractionAtoms,Muto1943ForceMolecules,Huang2015ReliableTheory}.

The damped ATM asymptotic dispersion model used in this work follows the workflow of Ref.~\citenum{Xie2023AssessmentTriazine} and expresses the three-body dispersion energy as a summation over triplets of atoms in the complex, one from each molecule\cite{Grimme2010AH-Pub,Xie2023AssessmentTriazine}: 
\begin{equation}
\label{eq:3B_ATM_Disp}
 E_{ATM} = \sum_{a\in A} \sum_{b\in B} \sum_{c\in C} f^{abc}_9 (\beta) E^{abc}_{ATM}
\end{equation}
where
\begin{equation}
\label{eq:triple-dipole3BATM}
  E^{abc}_{ATM} =  C^{abc}_9  \frac{ (1 + 3  \cos \theta _a \cos \theta _b \cos \theta _c)}{(R_{ab}R_{ac}R_{bc})^3}
\end{equation}
represents the triple-dipole ATM contribution in terms of interatomic distances $R_{ij}$ and angles $\theta_i$ in the triangle connecting the three atoms.
The appropriate atom-atom-atom dispersion coefficients $C^{abc}_9$ are modeled as $C^{abc}_9 \approx \sqrt{C^{ab}_6 C^{ac}_6 C^{bc}_6}$ using standard atom-atom dispersion coefficients $C^{ij}_6$ from the DFT-D4 formalism \cite{Caldeweyher2017ExtensionModel}. 
At short range, it is critical to damp the ATM dispersion terms to prevent unphysical divergence as well as account for overlap effects. The damping function $f^{abc}_9 (\beta)$ is chosen as a product of three two-body Tang–Toennies (TT) damping functions\cite{AnatoleVonLilienfeld2010Two-Solids, Tang1984AnCoefficients}.
\begin{equation}
\label{eqn:TT_Damping_fn}  
f^{abc}_9 (\beta) = f^{ab}_6 (R_{ab}\beta) f^{ac}_6 (R_{ac}\beta) f^{bc}_6 (R_{bc} \beta)
\end{equation}
\begin{equation}
f_6 (R\beta) = 1 - \sum _{k=0} ^6 \frac{(R\beta)^k}{k!}  e^{-\beta R}
\end{equation}
with the damping parameter $\beta$ estimated empirically as in Ref.~\citenum{Xie2023AssessmentTriazine}, which can be consulted for further details of the ATM algorithm employed here.

We performed the supermolecular benchmark and two-body SAPT(DFT) calculations using the MOLPRO program\cite{Werner2012Molpro:Package}. For all wavefunction-based SAPT calculations (from SAPT0 to SAPT2+3(CCD)$\delta$MP2), the Psi4 code was used\cite{Smith2020PChemistry}. The nonadditive SAPT(DFT) results were obtained using the three-body module of the SAPT package\cite{Bukowski2009Sapt2008Guide}, importing the DFT quantities from DALTON 2.0\cite{Aidas2014TheSystem}. The XSAPT computations utilized the Q-Chem code\cite{Shao2015AdvancesPackage}.
Finally, the ATM three-body dispersion terms were computed by a code written by Austin Wallace (Georgia Tech): see \verb+https://github.com/Awallace3/dispersion+.

\section{Results and Discussion}
\subsection{Benchmark interaction energies}
 
\begin{figure}
    \centering
    \includegraphics[width=\linewidth]{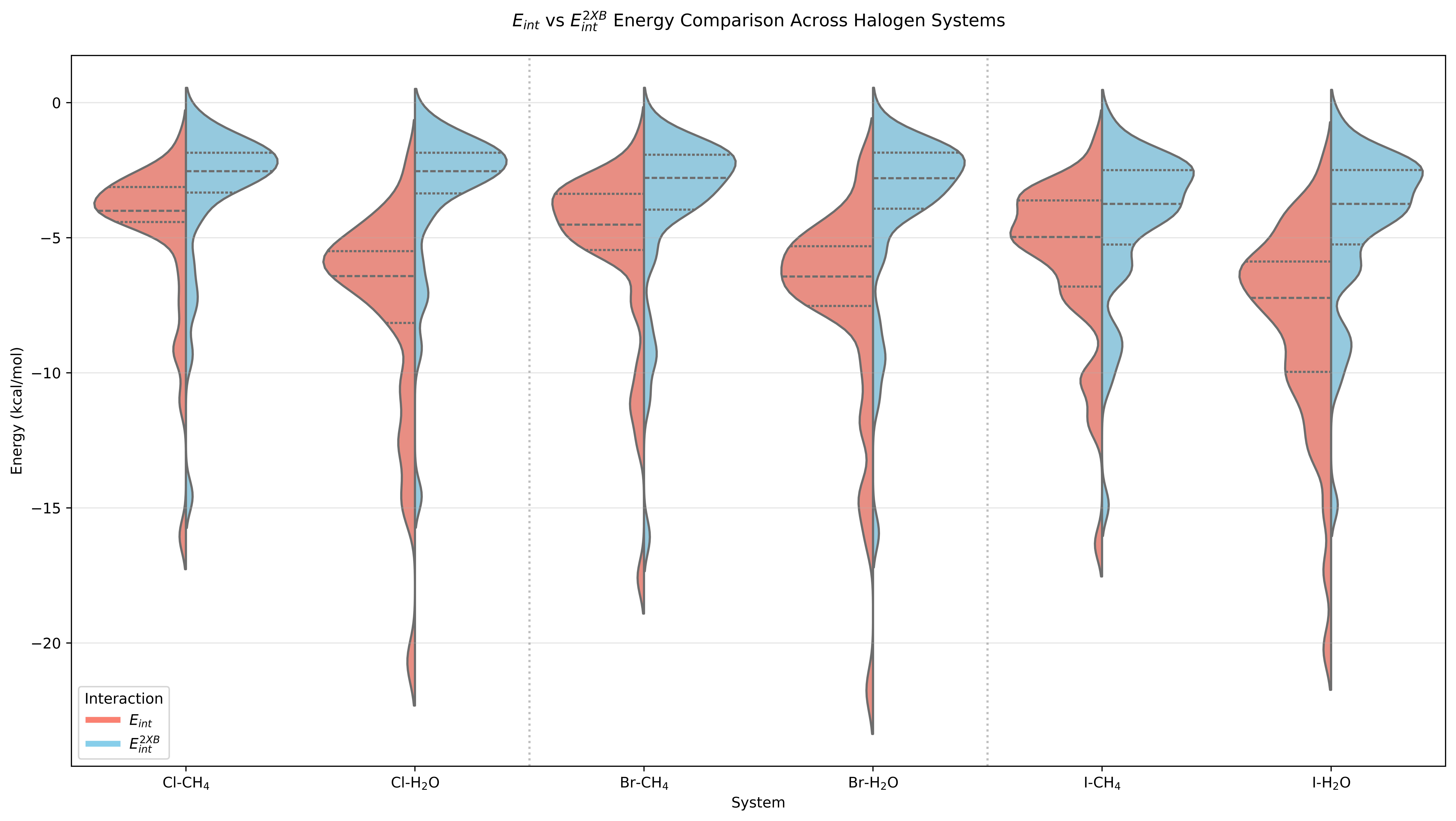}
    \caption{The distribution of complete interaction energies $E_{int}$ for the 3BXB complexes, partitioned by the halogen and by the third interaction partner (H$_2$O/CH$_4$), compared to the respective halogen-bonded value $E^{2XB}_{int}$, that is, the interaction energy for the unaltered halogen-bonded dimer AB. 
    }
    \label{fig:E3vsE2XB_int_benchmark}
\end{figure}

\begin{figure}
    \centering
    \includegraphics[width=\linewidth]{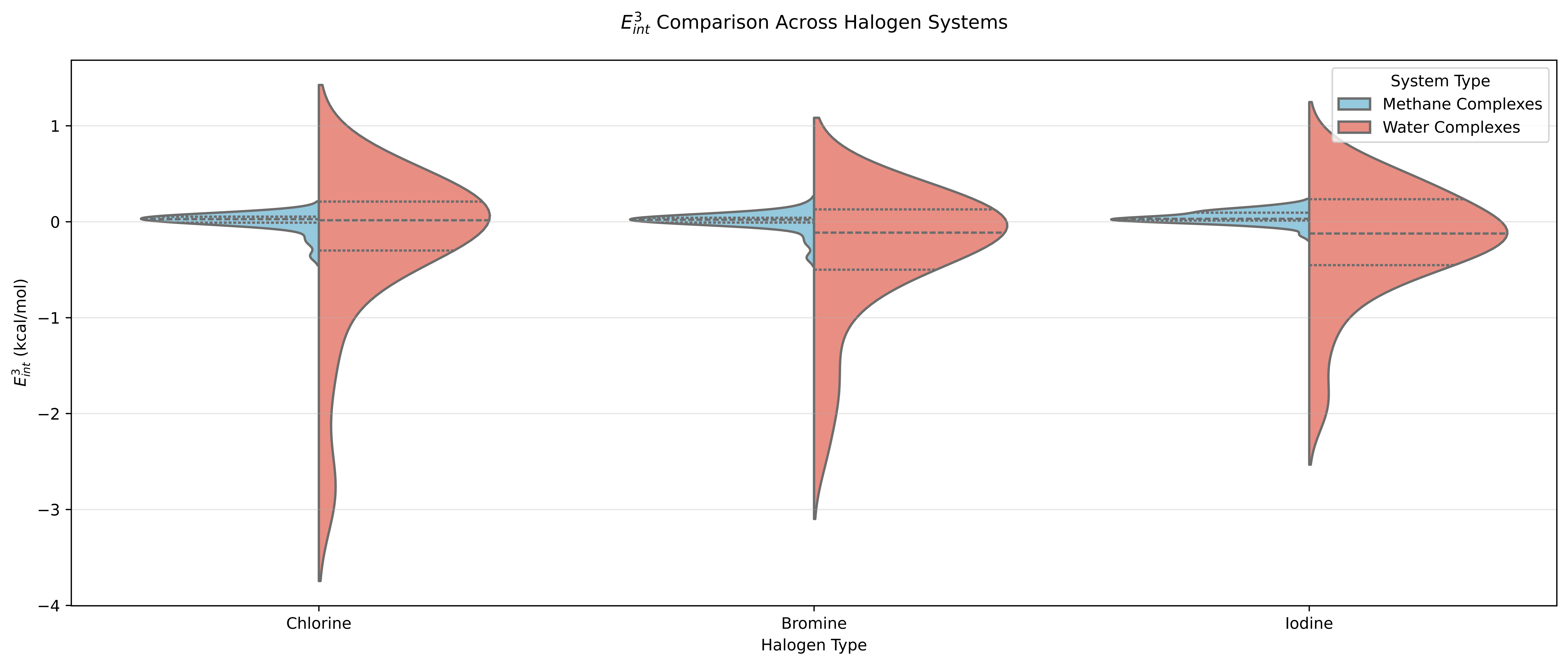}
    \caption{The distribution of the nonadditive three-body energies $E^3_{int}$ for the 3BXB complexes, partitioned by the halogen and by the third interaction partner (H$_2$O/CH$_4$).}
    \label{fig:Nonadd_int_benchmark}
\end{figure}

Figures \ref{fig:E3vsE2XB_int_benchmark} and \ref{fig:Nonadd_int_benchmark} provide a summary distribution of the total benchmark interaction energies as well as their nonadditive three-body contributions.
The individual benchmark values for each system are provided in Tables SIV--SIX in the Supporting Information.
The reference total interaction energy ranges from $-2.19$ to $-21.76$ kcal mol$^{-1}$ for water complexes and from $-1.48$ to $-17.59$ kcal mol$^{-1}$ for methane complexes.
The trimers with water show overall stronger binding (average interaction energy of $-7.84$ kcal mol$^{-1}$) than the trimers with methane (average interaction energy of $-5.38$ kcal mol$^{-1}$), 
with I showing the largest interactions; the average trimer interaction energy amounts to $-8.20$ kcal mol$^{-1}$ for iodine-water complexes compared to $-7.53$ kcal mol$^{-1}$ for bromine-water and $-7.69$ kcal mol$^{-1}$ for chlorine-water ones. This can be linked to the size and polarizability of the halogen atom involved in the interaction.
The benchmark nonadditive contribution can be either attractive or repulsive and ranges between $-2.87-+0.69$ kcal mol$^{-1}$ for trimers with water, and only between $-0.37-+0.18$ kcal mol$^{-1}$ for trimers with methane. Complexes 5  and 10 in Fig. \ref{fig:struct1} show the systems that exhibit the largest nonadditive effects for methane and water complexes, respectively, in the 3BXB dataset.
The relative magnitudes of the nonadditive interaction energies are in the range 0.4--23$\%$. 

In the next subsections, we will analyze the performance of various SAPT flavors in addition to the MP2 method applied to our set of $\sigma$-hole noncovalent interactions.  
The MP2/CBS data includes the correlation energy extrapolated to the CBS limit from aQZ and a5Z. The wavefunction SAPT (SAPT0 to SAPT2+3(CCD)$\delta$MP2), two-body SAPT(DFT), and XSAPT calculations were performed in the aTZ basis. 
The three-body SAPT(DFT) calculations were only feasible for complexes not involving iodine, and were performed in jun-cc-pVDZ for complexes with more than 25 atoms and aDZ otherwise. 
The XSAPT calculations are also problematic for iodine complexes (because of the need of an explicit-core-potential treatment of inner electrons), but we were able to perform incomplete XSAPT to extract the MBD three-body dispersion data.
We will first investigate the individual interaction energy contributions, at the two-body and three-body levels, before moving on to examining the accuracy of the entire interaction energies using mean unsigned errors (MUE) relative to the CCSD(T)-level benchmark values.

\subsection{SAPT two-body interaction energy contributions}

\begin{figure}
    \centering
    \includegraphics[width=\linewidth]{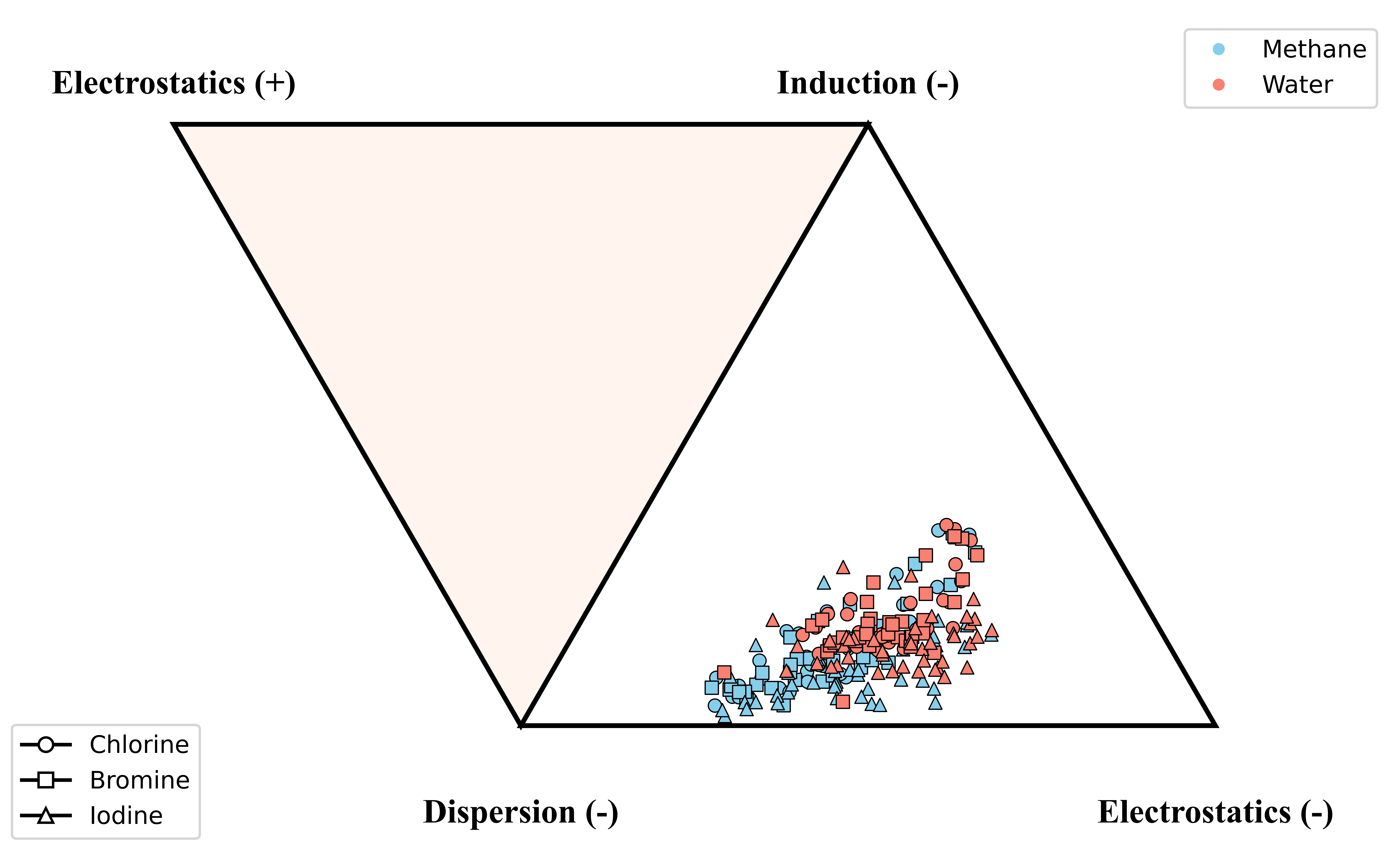}
    \caption{Two-body SAPT2+(CCD)$\delta$MP2 interaction energy distribution for different parts of the 3BXB dataset (separated by the XB halogen as well as water/methane), displayed on a ternary diagram.}
    \label{fig:f18}
\end{figure}

Tables SX--SXXV in the Supporting Information show the two-body interaction energy components for the 3BXB structures computed using SAPT2+3(CCD)$\delta$MP2, SAPT(DFT), and XSAPT. The electrostatic, induction, and dispersion components to the interaction energy are all attractive for those minimum structures. 
The magnitude of the interaction energy components depends on the halogen type with additional variation influenced by the presence of water or methane in the complex\cite{Rezac2012BenchmarkMolecules}.
The trimers constructed by adding water to the pre-existing dimers are predominantly bound by electrostatics with the  exception of the aromatic complexes that show larger dispersion magnitudes. Second-order dispersion is the leading term for complexes constructed with methane, with a few exceptions for systems with Cl$_2$, Br$_2$, and  I$_2$ as the donor.  
This behavior is expected because H$_2$O is a highly polar molecule, therefore significantly contributing to electrostatics whereas a nonpolar methane molecule drives dispersion forces. 
On the other hand, complexes with Cl$_2$, Br$_2$, and  I$_2$ (for example, structures 11 and 12 in Figure \ref{fig:struct1})  exhibit strong electrostatic interactions because of the increased polarizability of the halogen atom\cite{Rezac2012BenchmarkMolecules}. With an increase in size, and thus polarizability, of the halogen atom, the largest contributors to the interaction, namely electrostatics and dispersion, also get stronger. In addition, the presence of electron withdrawing groups enlarges the $\sigma$-hole by pulling electron density away from the halogen, strengthening the interaction, while electron-donating groups act in the opposite direction\cite{Rezac2012BenchmarkMolecules}.
A detailed SAPT2+(CCD)$\delta$MP2 energy decomposition analysis of the 3BXB dataset shows that the interactions span a relatively small section of the ternary diagram with a varying but fairly equal mix between dispersion and electrostatics   (Fig.~\ref{fig:f18}). 
The two-body energy decomposition highlights that the methane complexes cluster towards the dispersion dominated region while the water complexes towards the electrostatic corner, also with a stronger induction influence. 
The chlorine and bromine complexes tend to exhibit similar bonding, clustering in the same areas of the ternary diagram. On the other hand, many but not all iodine complexes cluster lower on the triangle, displaying somewhat diminished induction contributions.

As we will show in Sec.~\ref{sec:totalsapt}, the two-body SAPT2+(CCD)$\delta$MP2 level of theory recovers very well, and better than any other SAPT variant, the benchmark CCSD(T) pairwise additive interaction energies. Therefore, we will examine the interaction energy component differences from various SAPT flavors treating the SAPT2+(CCD)$\delta$MP2 values as reference.
Tables SXXXVII--SXL in the Supporting Information show the mean signed and unsigned deviations of each SAPT component between a given variant and the reference one.
We see that the simplest SAPT0 variant generally overbinds (the interaction energies are too negative), primarily because the simple $E^{(10)}_{exch}$ expression underestimates the reference first-order exchange energy (on the average by 1.72 kcal mol$^{-1}$ for water complexes and 1.26 kcal mol$^{-1}$ for methane complexes).
In contrast, XSAPT significantly underbinds, and the primary reason is underestimating the magnitude of the dispersion energy, on the average by 2.11 kcal mol$^{-1}$ for either water or methane complexes (this particular value is averaged over non-iodine systems only).
It seems like the short-range adjustment of the MBD formalism to adapt to XSAPT \cite{Carter-Fenk2019AccurateDispersionc} might benefit from a reoptimization for halogen-bonded systems.
The SAPT(DFT) dispersion energy is also underestimated in magnitude, albeit by much less (by 1.12 and 0.99 kcal mol$^{-1}$ on the average for trimers with water and methane, respectively) and again this underestimation is the leading cause of the SAPT(DFT) underbinding.
The correlated wavefunction-based SAPT variants exhibit smaller differences relative to one another, and these differences are harder to interpret. Note that the reference electrostatic energy is formally more approximated than the SAPT2+(3) and SAPT2+3 one because it omits the $E^{(13)}_{elst,resp}$ term, and the $E^{(30)}_{ind-disp}+E^{(30)}_{exch-ind-disp}$ contribution is computed explicitly and classified as dispersion in SAPT2+3$\delta$MP2, but forms a part of the $\delta$MP2 term, classified as induction in SAPT2+(3)$\delta$MP2\cite{Parker2014LevelsEnergies}.
The inclusion of the CCD+ST(CCD) dispersion terms beyond $E^{(20)}_{disp}+E^{(21)}_{disp}+E^{(22)}_{disp}$ makes the dispersion energy slightly less negative, by about 0.3 kcal mol$^{-1}$ on the average for either class of trimers.
Such a change is similar in magnitude to the differences in other SAPT corrections and may be beneficial or detrimental, however, it does strongly benefit the accuracy at the reference SAPT2+(CCD)$\delta$MP2 level.


\subsection{SAPT analysis of the three-body nonadditive interaction}

\begin{figure}
    \centering
    \includegraphics[width=\linewidth]{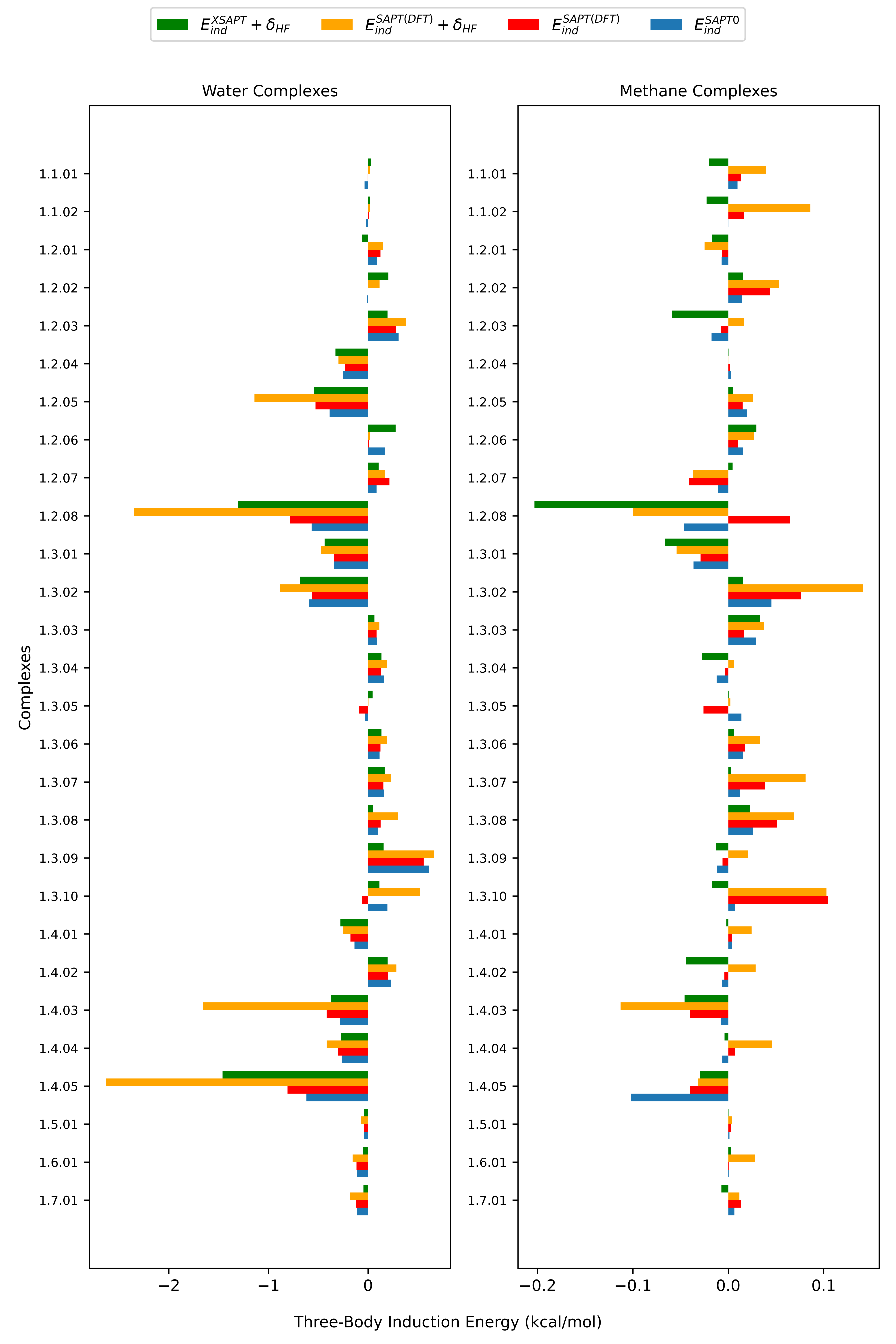}
    \caption{Comparison of the nonadditive three-body induction energies for the systems with chlorine as the XB donor, predicted by SAPT0, XSAPT, and SAPT(DFT). The system labels are defined in the Supporting information; the first `1.' part denotes chlorine. Note the different horizontal scale in the left and right panels. 
    }
    \label{fig:threeb_ind_cl}
\end{figure}

\begin{figure}
    \centering
    \includegraphics[width=\linewidth]{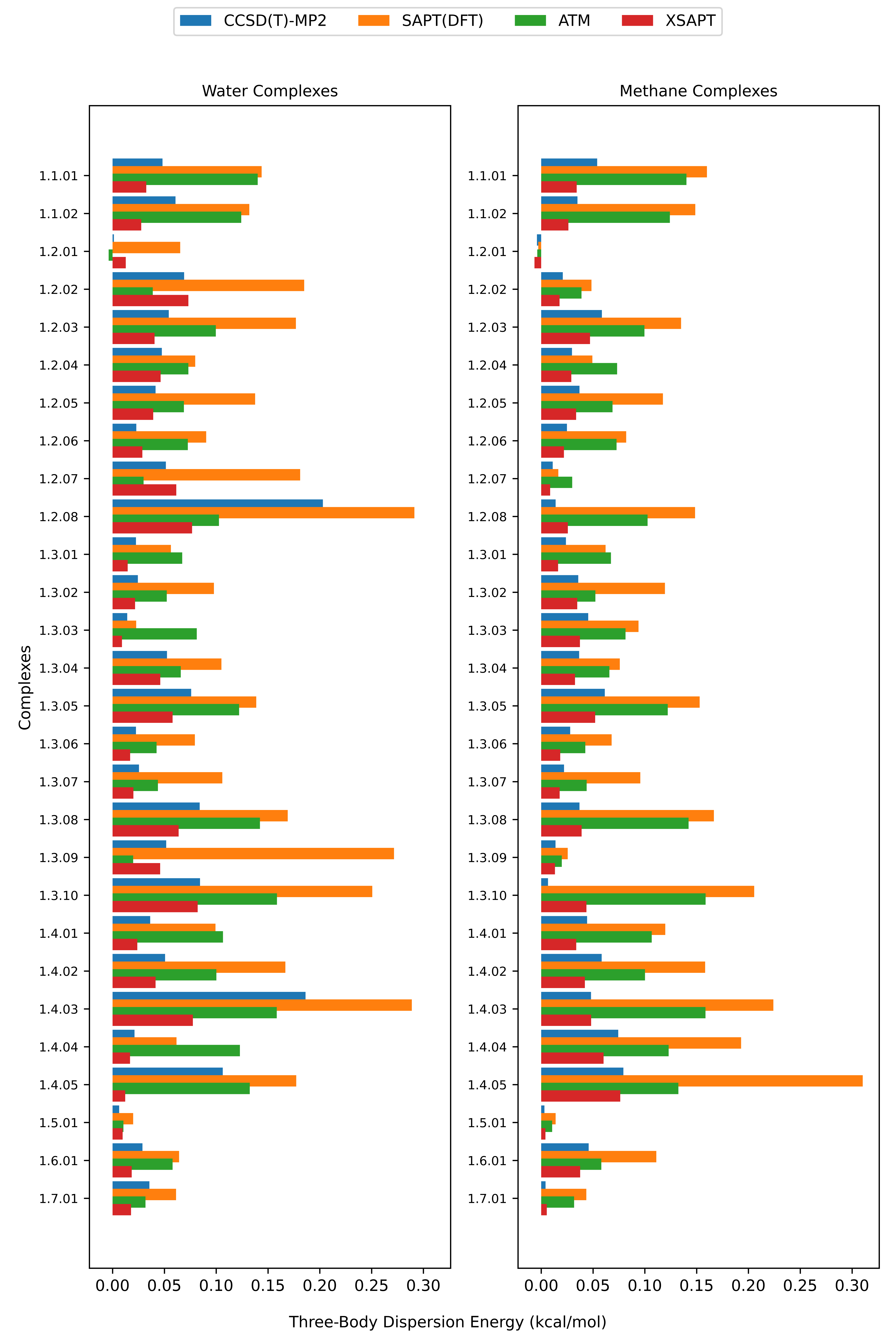}
    \caption{Nonadditive three-body dispersion energy for the systems with chlorine as the XB donor, estimated as the supermolecular CCSD(T)/CBS$-$MP2/CBS difference as well as computed with SAPT(DFT), XSAPT (that is, MBD) and estimated through the asymptotic ATM term. The system labels are defined in the Supporting information; the first `1.' part denotes chlorine.} 
    \label{fig:threeb_disp_cl}
\end{figure}

The SAPT energy decomposition of the nonadditive interaction in the 3BXB dataset shows that the three-body induction dominates while the three-body exchange and dispersion play smaller roles.  
Figures \ref{fig:Nonadd_int_benchmark}, \ref{fig:threeb_ind_cl} and  \ref{fig:threeb_disp_cl} present an analysis of the nonadditive interactions. Here, we can no longer use the two-body SAPT2+3(CCD)$\delta$MP2 approach, but instead shift our attention to SAPT(DFT) and XSAPT. 
As mentioned before, these nonadditive interaction energy calculations were not feasible for iodine complexes. Consequently, Figures~\ref{fig:threeb_ind_cl} and  \ref{fig:threeb_disp_cl} include data only from the chlorine complexes, and we present analogous figures for the bromine complexes (and partial nonadditive dispersion data for the iodine ones) in the Supporting Information. 
As explained above, SAPT(DFT) includes all the leading order nonadditive effects while XSAPT neglects the nonadditive first-order exchange component. 
XSAPT includes, however, nonadditive induction applicable to any number of molecules (not just three) and can be used with or without the three-body induction couplings\cite{Herbert2012RapidTheory}. 
As our earlier study \cite{Ochieng2023AccurateDecomposition} indicates that the latter couplings play a minor role, we neglect them here.

We now examine the SAPT(DFT) energy decomposition for the (non-iodine) 3BXB dataset, with the individual components for each system given in Tables SXXVI--SXXIX in the Supporting Information.
The nonadditive first-order exchange is predominantly attractive (negative for 112 systems and positive for 12 ones) and ranges from $-0.49$ to $0.20$ kcal/mol for the water complexes and from $-0.31$ to $0.003$ kcal/mol for the methane ones.
The nonadditive dispersion in this dataset is repulsive, aiming to remedy the overbinding that might come from the two-body dispersion contribution. 
This component at the SAPT(DFT) level contributes up to 0.33 kcal mol$^{-1}$ for complexes with water and up to 0.31 kcal mol$^{-1}$ for complexes with methane.
Conversely, the nonadditive induction can  be either attractive or repulsive, and it ranges from $-2.63$ to 0.66 kcal mol$^{-1}$ for trimers involving water and from $-0.24$ to 0.14 kcal mol$^{-1}$ for for trimers involving methane. 
An in-depth analysis of the nonadditive induction is shown in Fig. \ref{fig:threeb_ind_cl}. Here, we compare this three-body component computed by SAPT(DFT), SAPT0, and XSAPT. 
The 
SAPT(DFT) variant can include or omit the nonadditive $\delta E^{HF}$, while the $\delta E^{HF}$ contribution in XSAPT is assumed to be pairwise additive and thus has no effect on the nonadditive induction. The systems in the 3BXB dataset follow trends consistent with our previous work on heteromolecular trimers\cite{Ochieng2023AccurateDecomposition}, where we observed that SAPT0 and SAPT(DFT) without the $\delta E^{HF}$ correction have similar values of nonadditive induction. The incorporation of the $\delta E^{HF}$ term into SAPT(DFT) results in an increase in magnitude making the SAPT(DFT)+$\delta E^{HF}$ values the largest, followed by XSAPT. Overall, we do expect the inclusion of $\delta E^{HF}$ to improve accuracy as this term covers physical effects of nonadditive higher-order induction.  

Fig. \ref{fig:threeb_disp_cl} shows the different dispersion estimates from the supermolecular CCSD(T)$-$MP2, XSAPT, SAPT(DFT), and ATM approaches. 
Note that while supermolecular MP2 entirely misses the three-body dispersion, the CCSD(T)$-$MP2 difference is not strictly due to dispersion energy as the molecular multipole moments, polarizabilities, and electron densities somewhat change between MP2 and CCSD(T), leading to differences in other contributions.
Nevertheless, this difference has been previously taken as a measure of the three-body dispersion \cite{Kennedy2014Communication:Theory}. Consistent with previous observations from Huang and Beran\cite{Huang2015ReliableTheory}, Carter-Fenk et al.\cite{Carter-Fenk2021PredictingTheory,Carter-Fenk2019AccurateDispersionc}, and our earlier work \cite{Ochieng2023AccurateDecomposition},
the SAPT(DFT) and ATM models yield the largest dispersion estimates. 
While both of these methods are able to capture the repulsiveness of the many-body dispersion, they tend to overestimate it as they both neglect higher-order effects that are again attractive\cite{Carter-Fenk2019AccurateDispersionc,Huang2015ReliableTheory}. 
Indeed, relative to the leading asymptotic term (that is, ATM), SAPT(DFT) amplifies the nonadditive dispersion (through the inclusion of repulsive third-order effects beyond the triple dipole approximation) while XSAPT (that is, MBD) reduces it (through an effective inclusion of some attractive higher-order terms).
The MBD estimate, which is obtained from the supermolecular trimer–dimer differences, has the smallest magnitude and is mostly consistent with CCSD(T)$-$MP2, especially for the trimers involving methane. The reasonable agreement between XSAPT (MBD) and  CCSD(T)$-$MP2 suggests that both methods can reliably capture nonadditive dispersion in our systems. 
The MBD approach has been previously observed to produce better three-body interaction energies when combined with supermolecular MP2 than on top of supermolecular DFT since, in the latter case, MBD would have to account for deficiencies in the underlying density functional, which is not possible with a single damping/switching parameter \cite{Jankiewicz2018Dispersion-CorrectedEnergies}. In particular, unlike MP2, supermolecular DFT is not entirely free of three-body dispersion, and it has trouble correctly reproducing the nonadditive first-order exchange \cite{Hapka2017TheEffects}.
Those issues do not apply to either XSAPT or SAPT(DFT), so MBD has the potential to perform well in our case as long as the damping parameter value is suitable.

\subsection{Total SAPT interaction energies}
\label{sec:totalsapt}

So far, we have investigated the physical contributions to the two-body and three-body interaction energies in the 3BXB trimers as provided by various variants of SAPT.
As we want to exploit the full strength of SAPT, that is, the combination of a meaningful energy decomposition and accurate total interaction energies, we now examine the accuracy of complete interaction energies from different SAPT variants relative to our CCSD(T)-level benchmark data.
The interaction energy in a cluster is dominated by its pairwise-additive term, so we first assess the accuracy of two-body SAPT interaction energies.
Subsequently, we combine the best two-body SAPT performers with various measures of the three-body nonadditive contribution to identify the variants that most closely reproduce the reference total energies in the 3BXB trimers.
In this analysis, we do include hybrid SAPT approaches, in which the two-body and three-body contributions are obtained from different SAPT flavors. We even allow three-body estimates obtained by combining different sources, for example, the SAPT(DFT) nonadditive induction and first-order exchange with the MBD nonadditive dispersion.

We consider a large variety of SAPT variants with different intramolecular electron correlation treatments. 
These include SAPT0, where dispersion is introduced through second-order perturbation theory and the monomers are treated at the HF level (no intramolecular correlation), along with its progressive improvements through SAPT2, SAPT2+, SAPT2+(3), and SAPT2+3.
In the higher-level wavefunction SAPT variants (SAPT2+ and above), we also explore including additional corrections to dispersion (through the CCD+ST(CCD) level) and induction (the $\delta$MP2 term).
While these wavefunction-based SAPT treatments are only available for the two-body interaction energy, we also consider the SAPT(DFT) and XSAPT approaches which provide both the two-body and three-body decomposition.  

\begin{figure*}
    \centering
    \includegraphics[width=\textwidth]{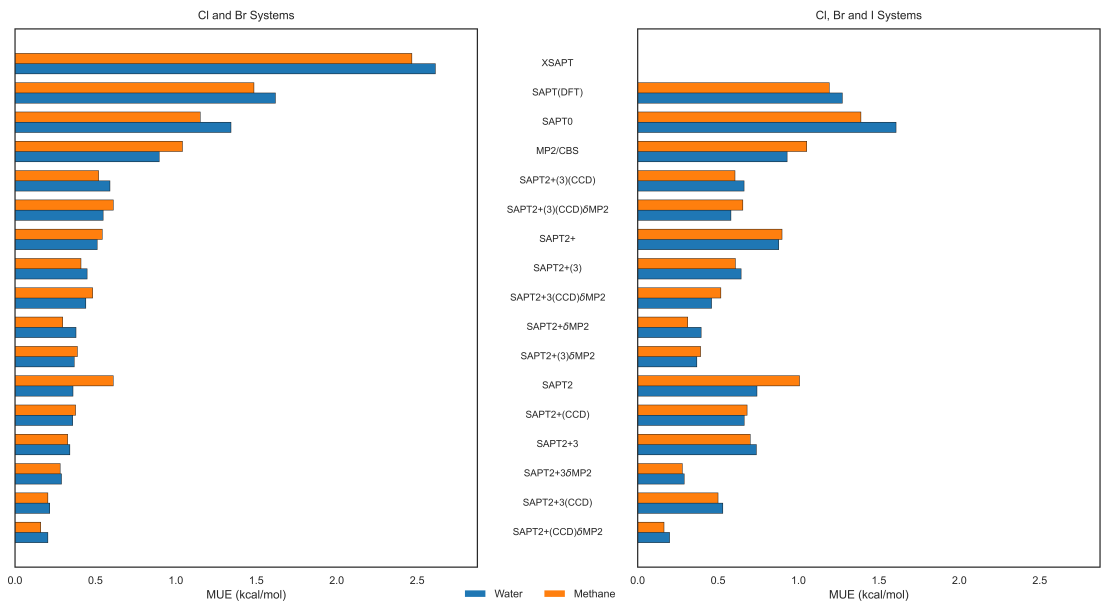}
    \caption{Mean unsigned errors (MUE, kcal mol$^{-1}$) of various two-body interaction energy estimates relative to the benchmark CCSD(T)-level $E^2_{int}$ values. The values in the left panel are averaged over the 124 complexes containing chlorine and bromine only, while the right panel values are averaged on the entire 214-system dataset.}
    \label{fig:2B_MUE}
\end{figure*}

\begin{figure}
    \centering
    \includegraphics[width=\linewidth]{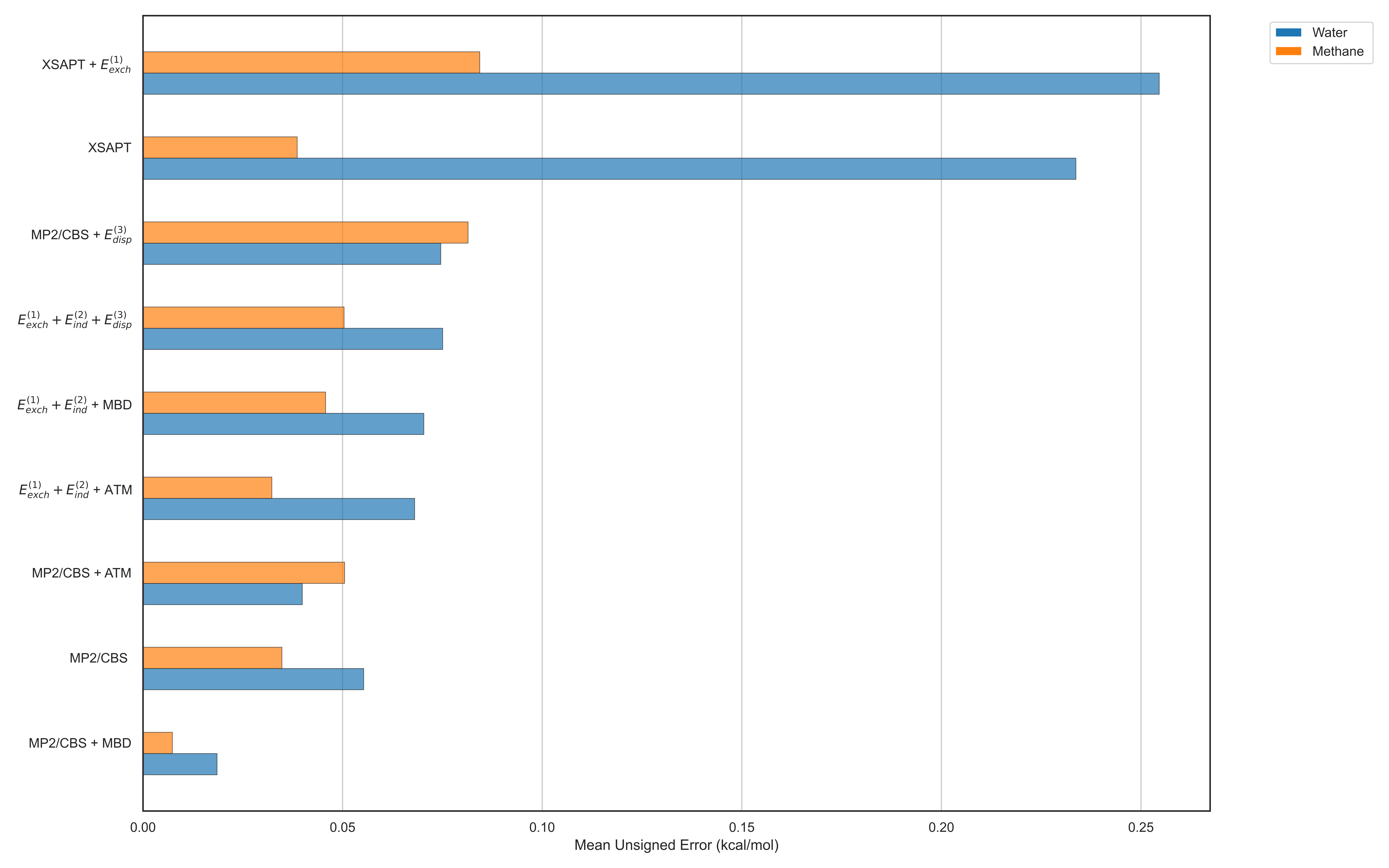}
    \caption{Mean unsigned errors (MUE, kcal mol$^{-1}$) of various nonadditive three-body interaction energy estimates relative to the benchmark CCSD(T)-level $E^3_{int}$ values. The corrections $E^{(1)}_{exch}$, $E^{(2)}_{ind}$, and $E^{(3)}_{disp}$ are taken from three-body SAPT(DFT), and their sum $E^{(1)}_{exch}+E^{(2)}_{ind}+E^{(3)}_{disp}$ is by definition the SAPT(DFT) three-body interaction energy.}
    \label{fig:Nonadd_MUE}
\end{figure}

\begin{figure*}
    \centering
    \includegraphics[width=\textwidth]{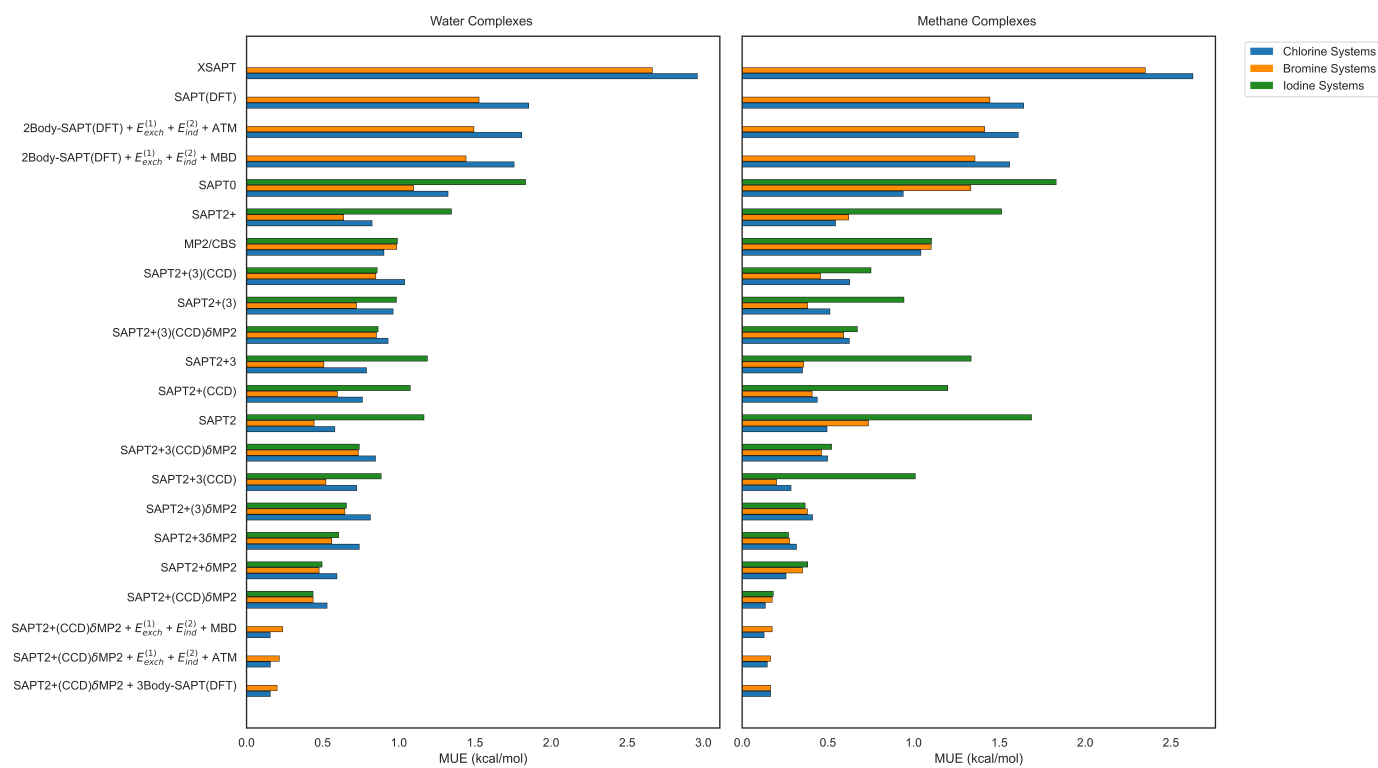}
    \caption{Mean unsigned errors (MUE, kcal mol$^{-1}$) of the total (two-body plus three-body) interaction energy estimates relative to the benchmark CCSD(T)-level complete $E_{int}$ values, grouped by the identity of the halogen bond donor. For consistency between different methods, the data with chlorine and bromine as the XB donor do not contain the 6 systems (3 with water and 3 with methane) involving iodine in a different capacity (for which the XSAPT and three-body SAPT(DFT) data are not available).}
    \label{fig:3B_MUE}
\end{figure*}

The mean unsigned errors of different approaches relative to the reference CCSD(T)-level data are presented in Figs.~\ref{fig:2B_MUE}, \ref{fig:Nonadd_MUE}, and \ref{fig:3B_MUE} for the two-body, nonadditive three-body, and total trimer interaction energies, respectively.
Since our perturbative calculations of three-body contributions, both SAPT(DFT) and XSAPT, are limited to complexes not involving iodine, the errors in Figs.~\ref{fig:Nonadd_MUE} and \ref{fig:3B_MUE} are averaged over the 124 non-iodine complexes (62 each with water and methane).
In the two-body case, the analogous MUE values in Figs.~\ref{fig:2B_MUE} are given both for the 124 non-iodine systems (left panel) and for the full set of 214 complexes (right panel).

The ordering of the two-body errors presented in Fig.~\ref{fig:2B_MUE} is overall quite expected. The lowest SAPT0 level of theory leads to the largest MUE value on the whole dataset, although XSAPT yields the highest errors for the reduced dataset for which it was computed).
Interestingly, the SAPT(DFT) variant is overall as inaccurate as SAPT0, and both SAPT0 and SAPT(DFT) exhibit comparable errors for the water and methane trimers, indicating that the error of the sum $E^{AB}_{int}+E^{BC}_{int}+E^{CA}_{int}$ is dominated by the (common to both) halogen-bonded dimer $E^{AB}_{int}$.
All correlated levels of wavefunction-based SAPT, starting from SAPT2, are more accurate than SAPT(DFT), and they also all surpass the accuracy of supermolecular MP2. Among these correlated variants, those that include the $\delta$MP2 correction are clearly superior, indicating that, without this term, the SAPT data are missing some important induction-dispersion couplings or some exchange effects beyond the commonly used approximations.
The best performing SAPT variant happens to be SAPT2+(CCD)$\delta$MP2, as it clearly combines a high level of SAPT theory with a favorable cancellation of remaining residual errors. SAPT2+(CCD)$\delta$MP2 leads to impressively low MUE values of 0.20 (trimers with water) and 0.16 kcal mol$^{-1}$ (trimers with methane) when averaged over either the full 214-complex dataset or the 124-complex reduced non-iodine dataset.
It should be noted that many other variants of SAPT perform worse on the complexes containing iodine than for the lighter halogens, which leads to SAPT errors typically becoming smaller when evaluated over the reduced dataset.

We move on to the nonadditive three-body interaction energy contributions displayed in Fig.~\ref{fig:Nonadd_MUE}.
Here, our assortment of methods has shrunk since the correlated wavefunction-based SAPT approaches are only available in the two-body case. 
In the context of the energy decomposition presented above, only SAPT(DFT) can account for all leading sources of three-body nonadditivity, while XSAPT is missing first-order nonadditive exchange and supermolecular MP2 is missing nonadditive dispersion.
On the other hand, three-body SAPT(DFT) has been observed, here and earlier \cite{Podeszwa2007Three-bodyMonomers}, to overestimate nonadditive dispersion, so it is worthwhile to examine the performance of hybrid SAPT(DFT)-based variants replacing its CKS dispersion by either MBD (as included in XSAPT) or the leading asymptotic term, ATM. The same measures of dispersion can be added to supermolecular MP2.

The errors of different nonadditive three-body estimates in Fig.~\ref{fig:Nonadd_MUE} are overall quite similar to each other, with an exception of XSAPT which is less accurate for the trimers with water.
At first, one could think that this is due to XSAPT neglecting nonadditive first-order exchange. However, the addition of the SAPT(DFT) $E^{(1)}_{exch}$ estimate to XSAPT makes the errors larger, not smaller, leading us to believe that the issue with XSAPT might be its comparatively inaccurate description of nonadditive induction.
The MUE values of all other methods do not exceed 0.1 kcal mol$^{-1}$ for either interaction partner (water or methane), which is significantly lower than the two-body errors, indicating that any of those variants is in principle an acceptable strategy to account for nonadditive effects in our trimers.
The complete SAPT(DFT) nonadditive energy has a very reasonable average error of 0.075 and 0.050 kcal mol$^{-1}$ for the trimers with water and methane, respectively.
Supermolecular nonadditive MP2 performs slightly better (with MUE of 0.055 and 0.035 kcal mol$^{-1}$, respectively, for the water and methane trimers), and further improvement to the accuracy can be attained by hybrid nonadditive approaches. The SAPT(DFT) nonadditive exchange and induction performs best when combined with the ATM three-body dispersion estimate (with water and methane MUE values of 0.068 and 0.032 kcal mol$^{-1}$, respectively) and it can also be combined with the MBD three-body dispersion for a very similar accuracy.
The closest recovery of the benchmark nonadditive energies, with the water and methane MUE amounting to a respective 0.019 and 0.007 kcal mol$^{-1}$, is afforded by combining the supermolecular MP2 method, which accounts for the exchange and induction nonadditivity, with the three-body dispersion estimate from MBD.
While such a good agreement has to be to some extent accidental, it does lend even more credibility to MBD as the truthful descriptor of nonadditive dispersion effects in our halogen-bonded clusters.

We now analyze how different combinations of two-body and three-body approaches influence the accuracy of total interaction energies in the 3BXB clusters, employing the MUE values for supermolecular, SAPT, and hybrid approaches displayed in Fig.~\ref{fig:3B_MUE}.
As expected, the two-body accuracy is the most important factor, and the best two-body performer, SAPT2+(CCD)$\delta$MP2, leads to reasonable MUE values of 0.48 and 0.16 kcal mol$^{-1}$ for water and methane trimers, respectively (computed on the 124-member reduced non-iodine dataset), even without any nonadditive three-body estimate.
This level of accuracy is not attained by any available method that provides both two-body and three-body interaction energies, including SAPT(DFT), XSAPT, and supermolecular MP2. Only hybrid approaches can further improve the accuracy of the total interaction energies, and it is no surprise that the best hybrid variants take the best two-body performer, SAPT2+(CCD)$\delta$MP2, and augment it with an estimate of the nonadditive three-body effects.
For the latter estimate, all choices (including a choice not to add anything at all) perform equally well for the methane-containing trimers, with all MUE values in the 0.15--0.17 kcal mol$^{-1}$ range. For the water-containing trimers, there is a bit more variability, and the lowest MUE of 0.18 kcal mol$^{-1}$ is afforded by the addition of the complete SAPT(DFT) three-body energy to the SAPT2+(CCD)$\delta$MP2 value.
However, nearly the same accuracy is obtained if one keeps the SAPT(DFT) nonadditive induction and first-order exchange, but replaces its nonadditive dispersion by either MBD or ATM.
Overall, the best strategy to compute both accurate trimer interaction energies and their decomposition into physically meaningful terms appears to be combining a correlated wavefunction-based two-body SAPT variant utilizing the $\delta$MP2 correction with any SAPT(DFT)-like decomposition of the three-body energy, with dispersion either taken from SAPT(DFT) itself, computed with MBD, or approximated by its leading ATM term.

For the two methods that show the worst total interaction energy accuracy, XSAPT and SAPT(DFT), the biggest outliers are the chlorine$-$trimethylamine$-$water/methane trimers for both methods. These systems show respective errors of 8.59 and 6.26 kcal mol$^{-1}$ for XSAPT and 6.80 and 6.17 kcal mol$^{-1}$ for SAPT(DFT).
For the two-body interaction energy only, the worst errors pertain to XSAPT with the   chlorine$-$trimethylamine$-$water/methane complexes (7.09 and 6.08 kcal mol$^{-1}$, respectively), SAPT(DFT) also with chlorine$-$trimethylamine$-$water/methane (6.32 and 5.90 kcal mol$^{-1}$, respectively),
and SAPT0 for the iodine$-$pyridine-N-oxide$-$water/methane systems ($-7.16$ and $-5.18$ kcal mol$^{-1}$, respectively). 
Most methods in Fig.~\ref{fig:3B_MUE} perform very similarly for all XB donors, Cl, Br, and I (when the latter is available). One exception is the wavefunction-based SAPT without $\delta$MP2 being somewhat less accurate for iodine complexes (this difference is completely alleviated when $\delta$MP2 is added). Second, all SAPT variants seem to perform a little worse for chlorine-water systems than for bromine-water ones. However, the differences are not particularly significant.

\section{Conclusions}

In this work, we constructed a new 3BXB benchmark dataset which contains noncovalently interacting trimers with two halogen-bonded molecules taken from the SH250 dataset \cite{Kriz2022Non-covalentInteractions} and either water or methane as the third partner.
The benchmark interaction energies and their nonadditive three-body contributions were generated at the composite CCSD(T) level of theory, with basis sets appropriate for a ``silver-standard'' calculation. 
The new 214-member dataset helps alleviate the scarcity of benchmark-quality nonadditive interaction data, especially for trimers composed of three different molecules, where our previous 20-member dataset \cite{Ochieng2023AccurateDecomposition} was, to our knowledge, the only one of its kind.

The decomposition of both two- and three-body interaction energies into physically meaningful terms provides key insights into the nature of specific interactions.
Therefore, we used several variants of SAPT to analyze the physical origins and cooperativity of the binding in the 3BXB systems.
On the two-body level, the high-level wavefunction-based SAPT variants including the so-called $\delta$MP2 correction accurately reproduced the benchmark total interaction energies, and indicated that the 3BXB complexes are held together, in various proportions, by electrostatic and dispersion forces.
The nonadditive three-body interaction energy can in turn be decomposed into the effects of first-order exchange (mostly attractive), induction (dominant, can be attractive or repulsive), and dispersion (repulsive).  
While the first of those terms is only provided by SAPT (in this case, the three-body SAPT(DFT) variant of Ref.~\citenum{Podeszwa2007Three-bodyMonomers}), the other two terms can be computed from various sources. 
As far as nonadditive induction is concerned, the SAPT(DFT) and XSAPT values are reasonably consistent only when the $\delta$HF term is not included in the former; however, this term represents important high-order induction effects and should be normally taken into account.
The nonadditive dispersion energies, while small, display quite significant inconsistencies.
Relative to the leading-order asymptotic ATM contribution, the three-body SAPT(DFT) mostly predicts dispersion energy enhancement while MBD (as included in the XSAPT formalism) mostly suggests a reduction in magnitude for this effect.
The supermolecular CCSD(T)$-$MP2 interaction energy difference, which does include three-body dispersion but also contains other effects, also predicts reduced dispersion energies relative to ATM but exhibits some inconsistencies relative to MBD.
It appears that further analysis and benchmarking of the nonadditive dispersion effects is needed to provide a more comprehensive understanding.

Different SAPT variants display a broad range of accuracy relative to the reference CCSD(T)-level interaction energies.
Some high-level variants of wavefunction-based SAPT, especially SAPT2+(CCD)$\delta$MP2, are so accurate for the dominant two-body interaction energy term that they overperform SAPT(DFT) and XSAPT (variants that can be used for all two-body and three-body terms) even as they are, neglecting the nonadditive contributions entirely. 
However, the highest interaction energy accuracy is attained by hybrid approaches generated by mixing and matching different two-body and three-body estimates.
Overall, the best strategy that yields both accurate interaction energies and their physical decomposition involves augmenting a high-level wavefunction-based two-body SAPT energy such as SAPT2+(CCD)$\delta$MP2 with the SAPT(DFT) estimates of nonadditive induction and first-order exchange plus {\em any} estimate of nonadditive dispersion, even from the simple damped asymptotic ATM expression.

\section*{Conflicts of interest}

There are no conflicts to declare.

\section*{Data availability}

The numerical data computed in the course of this project are available in the article and the Supporting Information. The latter document also contains XYZ geometries of all studied complexes.

\section*{Acknowledgements}

This work was supported by the U.S. National Science Foundation
award CHE-1955328. We thank Austin Wallace (Georgia Tech) for the code to compute the ATM three-body contribution.



\balance

\renewcommand\refname{References}

\providecommand*{\mcitethebibliography}{\thebibliography}
\csname @ifundefined\endcsname{endmcitethebibliography}
{\let\endmcitethebibliography\endthebibliography}{}

\end{document}